\documentclass[usenatbib]{mnras}
\usepackage[T1]{fontenc}
\usepackage{times,ae,aecompl}
\usepackage{graphicx}	
\usepackage{amsmath}	
\usepackage{amssymb}	
\usepackage{wasysym}
\usepackage{xcolor}
\usepackage{multirow}
\usepackage{pdfcolfoot}
\usepackage[normalem]{ulem}
 \usepackage{url}
\usepackage{etoolbox}
\def\be{\begin{equation}}
\def\ee{\end{equation}}
\definecolor{bittersweet}{rgb}{1.0, 0.44, 0.37}
\definecolor{darkgreen}{rgb}{0.34, 0.55, 0.23}

\newcommand {\gev}{{\em gevolution}}
\newcommand {\MG}{{\texttt {MG-evolution}}}

\newcommand{\HH}{ \mathcal{H}}

\newcommand{\Geff}{G_{\rm eff}}
\newcommand{\rhom}{\rho_{\rm m}}
\newcommand{\brhom}{\bar{\rho}_{\rm m}} 
\newcommand{\deltarhom}{\delta\rho_{\rm m}}

\newcommand{\fzeta}{f_{\zeta}}
\newcommand{\rth}{r_{\rm th}}
\newcommand{\class}{\texttt{CLASS }}

\usepackage{graphicx}	
\usepackage{amsmath}	
\title{$N$-body simulations for parametrised modified gravity}

\author[Hassani \& Lombriser]{Farbod Hassani,$^{1}$\thanks{E-mail: Farbod.Hassani@unige.ch}
Lucas Lombriser,$^{1}$\thanks{E-mail: Lucas.Lombriser@unige.ch}
\\
$^{1}$D\'epartement de Physique Th\'eorique, Universit\'e de Gen\`eve, 24 quai Ernest Ansermet, 1211 Gen\`eve 4, Switzerland
\\}

\date{Accepted XXX. Received YYY; in original form ZZZ}
\pubyear{2019}

\begin{document}
\label{firstpage} 
\pagerange{\pageref{firstpage}--\pageref{lastpage}}
\maketitle 

\begin{abstract}
We present \MG, an $N$-body code simulating the cosmological structure formation for parametrised modifications of gravity.
It is built from the combination of parametrised linear theory with a parametrisation of the deeply nonlinear cosmological regime extrapolated from modified spherical collapse computations that cover the range of known screening mechanisms.
We test \MG, which runs at the speed of conventional $\Lambda$CDM simulations, against a suit of existing exact model-specific codes, encompassing linearised and chameleon $f(R)$ gravity as well as the normal branch of the Dvali-Gabadadz-Porrati braneworld model, hence covering both large-field value and large-derivative screening effects.
We compare the nonlinear power spectra produced by the parametrised and model-specific approaches over the full range of 
scales set by the box size and resolution of our simulations, $k=(0.05-2.5)$~h/Mpc, and for two redshift slices, $z=0$ and $z=1$.
We find sub-percent to one-percent level recovery of all the power spectra generated with the model-specific codes for the full range of scales.
\MG~can be used for generalised and accurate tests of gravity and dark energy with the increasing wealth of high-precision cosmological survey data becoming available over the next decade.
\end{abstract}

\begin{keywords}
large-scale structure -- $N$-body simulations -- modified gravity
\end{keywords}

\section{Introduction}

A wealth of high-precision measurements in the Solar System, of astrophysical objects, of gravitational wave emissions, and in the laboratory have put Einstein's Theory of General Relativity (GR) under intense scrutiny \citep{1975ApJ...195L..51H,Kapner:2006si,Will:2014kxa,TheLIGOScientific:2016src,Baker:2019gxo}.
In these regimes GR has successfully passed all tests so far.
Its application to cosmology, however, involves vastly different length scales, which in orders of magnitude compare to the extent of the Solar System as the scale of everyday human experience to the scale of an atomic nucleus.
It is therefore important to conduct independent tests of GR in the cosmological regime.
The necessity of a dominating dark sector to explain the cosmic large-scale observations \citep{1998AJ....116.1009R,1999ApJ...517..565P,Aghanim:2018eyx,Hildebrandt:2016iqg,Abbott:2017wau} 
 provides additional motivation for this endeavour.
Traditionally the late-time accelerated expansion of our Universe has been a particularly important driver for the development of modifications of GR.
As direct cause of the acceleration this motivation is, however, challenged by the confirmation of a luminal speed of gravity \citep{Lombriser:2016yzn,Monitor:2017mdv}.
But cosmic acceleration could nonetheless be attributed to a dark energy component that may couple nonminimally to matter, modifying gravity and leaving an observable impact on cosmological scales.

The past two decades have seen a steep growth in cosmological tests of gravity and dark energy~\citep{Koyama:2015vza,Joyce:2016vqv,Ishak:2018his},
and in the upcoming years, we will benefit from numerous high-precision experiments~\citep{Aghamousa:2016zmz,4MOST:2019,Laureijs:2011gra,Abate:2012za,Bull:2018lat} 
 that will enable us to put tight cosmological constraints on the properties of dark energy and modified gravity theories.
Specifically, we will see an increase in the wealth of high-quality data in  the nonlinear regime of cosmic structure formation.
This is a regime of particular interest for tests of gravity since viable modifications of GR must employ screening mechanisms \citep{Vainshtein:1972sx, Khoury:2003aq, Babichev:2009ee, Hinterbichler:2010es}
to recover GR in high-density regions to comply with the stringent bounds from Solar-System, astrophysical, gravitational wave, and laboratory experiments.
These are inherently nonlinear effects and naturally in the nonlinear cosmological small-scale structure is where modified gravity transitions to GR and thus also where unique signatures of screening are to be expected. 
The forthcoming nonlinear cosmological survey data therefore make cosmological tests of gravity a very timely enterprise.

To predict the complex nonlinear structure formation due to gravity, we typically rely on $N$-body simulations.
$N$-body codes have been developed for a range of alternative gravity models \citep{Oyaizu:2008tb,Schmidt:2008tn,Zhao:2010qy,2012JCAP...01..051L,Brax_2012,2012PDU.....1..162B,Puchwein:2013lza,Wyman:2013jaa,Barreira:2013eea,Li:2013tda,Llinares:2013jza,Mead:2015yca,Valogiannis:2016ane,Hassani:2019lmy} (for an introductory text see \cite{10.1088/978-0-7503-1587-6}).
Importantly, these simulations model the cosmic structure of specific modified gravity theories.
However, a plethora of modified gravity models are conceivable based on the prospects of novel interactions of matter with new fields.
A systematic approach is therefore required to more comprehensively explore the manifold cosmological implications from the possible modifications of gravity.
But this is not feasible with a model-by-model implementation in $N$-body codes.
Hence, much effort has gone into developing parametrisation frameworks (see \cite{Lombriser:2018guo} for a review).
While parametrisations of the linear and quasilinear modifications are well understood, it is less clear how to develop parametrisations of modified gravity in the deeply nonlinear cosmological regime.
In general, spherical collapse computations have proven very useful in capturing
the modified gravity effects at deeply nonlinear scales.
Motivated by the variety of screening mechanisms that can operate in scalar-tensor theories, a parametrisation of the modified gravitational forces acting on the spherical top-hat overdensities has been developed in \cite{Lombriser_2016}.
Together with linear and quasilinear parametrisations the spherical collapse parametrisation can be used in an adapted halo model framework to model the nonlinear matter power spectra of arbitrary modified gravity theories \citep{Cataneo:2018cic,Bose:2020wch}.
An $N$-body implementation of such a parametrisation framework, covering the linear to deeply nonlinear regime, has so far not been developed.
$N$-body codes for parametrised modified gravity exist in the context of large-field value screening \citep{Brax:2012gr}, which however does not encompass models with large-derivative screening, or for purely phenomenological parametrisations \citep{2011arXiv1112.6378T}, which however do not accurately represent the effect of screening mechanisms in the cosmic structure formation.
A physically motivated but general parametrisation of modified gravity effects for the implementation in $N$-body codes would both allow to simulate specific models with one code as well as to broadly parametrise and explore the modified gravity effects in simulations for the multitude of conceivable models and test these against the future survey data.
In this paper, we develop and present the first $N$-body code, dubbed \MG, for parametrised modifications of gravity that encompass all known screening effects.
We derive the parametrisation from generalised linear theory and the parametrised spherical collapse model of \cite{Lombriser_2016}.
We describe our parametrised simulations and test the performance for three types of modifications for which exact $N$-body simulation data are available. These are the linearised and chameleon $f(R)$ gravity models~\citep{Hu:2007nk} and the normal branch Dvali-Gabadadze-Porrati (nDGP) model~\citep{Dvali:2000hr}.
For a comparison of the simulation outputs we present the nonlinear matter power spectra produced by these models within the parametrised and exact approaches.

The paper is organised as follows.
In Sec.~\ref{sec:MG}, we review the linear and nonlinear parametrisations of modified gravity in Fourier and real space, respectively.
We then introduce a Fourier-space parametrisation that covers all scales and discuss its implementation in $N$-body codes.
We develop and test our \MG~$N$-body implementation in Sec.~\ref{sec:test}.
We test the performance of the parametrised code in reproducing the exact $N$-body results of existing codes for linearised and chameleon $f(R)$ gravity as well as the nDGP model.
We conclude with a discussion of our results in Sec.~\ref{sec:conclusions}.
Finally, we discuss details of moving beyond the quasistatic approximation and differences between simulations of parametrised modified gravity in Fourier and real space in the appendix.

\section{Parametrising modified gravity for $N$-body simulations} \label{sec:MG}

The cosmic structure formation in a given gravitational theory is most accurately modelled with $N$-body simulations, where the particles in the simulation are incrementally displaced from their initial positions according to the Poisson equation.
This computation is conveniently performed in Fourier space, where
the effective Poisson equation is given by
\begin{equation}
 \left(\frac{k}{a}\right)^2 \Phi (a,k) = 4\pi\Geff(a,k,\ldots) \deltarhom (a,k) \,. \label{eq:MGPoisson}
\end{equation}
$\Geff(a,k,\ldots)$ denotes the effective gravitational coupling, in general a function of time and wavenumber $k$, which parametrises the effects of alternative gravity theories on structure formation.
The gravitational potential $\Psi$ is cast in the Poisson gauge with the line element
\be
ds^2 = a^2(\tau) \Big[ - e^{2 \Psi} d\tau^2 + e^{-2 \Phi} \delta_{ij}   dx^i dx^j \Big]
\ee
and $\deltarhom$ denotes the matter density perturbation.

In addition to the modification of the Poisson equation, modified gravity models typically also introduce a gravitational slip between the spatial and temporal gravitational potentials, quantifying an effective anisotropic stress.
In general, modifications of gravity can also change the cosmological background expansion,
but our focus here is on the parametrisation of linear (Sec.~\ref{sec:lin}) and nonlinear (Sec.~\ref{sec:nl}) effects on structure formation.
For simplicity, for the practical examples in Sec.~\ref{sec:test} we shall therefore specify to models with a $\Lambda$CDM expansion history, but we stress that the formalism introduced here is not dependent on this choice.
For a review of general parametrisations of modified gravity, we refer the reader to~\cite{Lombriser:2018guo}.

\subsection{Linear parametrisation} \label{sec:lin}

At the level of linear perturbations and in the quasistatic regime, where we neglect time derivatives with respect to spatial derivatives, the effective modification of the Poisson equation of local four-dimensional metric theories of gravity that lead to at most second spatial derivatives in the equations of motion takes the form \citep{Silvestri_2013}
\begin{equation}
 G_{\rm eff, L}(a,k) = \frac{1+p_1(a)k^2}{p_2(a)+p_3(a)k^2} \,, \label{eq:GeffQS}
\end{equation}
where $p_i(a)$ are generally three independently free functions of time.
Note that more accurately one may adopt a semi-dynamical approximation~\citep{Lombriser_2015}, as discussed in App.~\ref{app:semi_analytics}, which includes the evaluation of the time derivatives at a pivot scale that can be absorbed into Eq.~(\ref{eq:GeffQS}).
This correction typically only contributes at near-horizon scales, but it becomes particularly important for scalar-tensor theories with higher-order derivatives, where it contributes to leading order at all linear scales.

\subsection{Nonlinear parametrisation} \label{sec:nl}

A parametrisation of modified gravity effects on the nonlinear cosmic structure formation that captures all known screening mechanisms has been proposed in \cite{Lombriser_2016} through the parametrisation of the spherical collapse equations in real space.
The effective gravitational coupling in this approach can be written as
\begin{equation}
 \frac{\Geff(a,r)}{G} = A + \sum_i^{N_0} B_i \prod_j^{N_i} \mathcal{F}_{ij} \,, \label{eq:nlparam}
\end{equation}
where $A$ corresponds to the effective coupling in the fully screened limit, typically unity, $B_i$ is the coupling in the fully unscreened limit, e.g., Eq.~\eqref{eq:GeffQS} that can be computed using the linear theory, $N_0$ and $N_i$ characterise the respective number of transitions, and $i,j$ are positive integers.
The $\mathcal{F}_{ij}$ are some transition functions parametrising screening or
other suppression effects.
To parametrise these transitions \cite{Lombriser_2016} adopted a generalised form of the Vainshtein screening effect in nDGP (Sec.~\ref{sec:nDGP}) with
\begin{equation}
 \mathcal{F} \sim b \left(\frac{r}{r_0}\right)^{a_f} \left\{ \left[1+\left(\frac{r_0}{r}\right)^{a_f}\right]^{1/b} - 1 \right\} \,, \label{eq:transition}
\end{equation}
where $r_0$ denotes the screening scale, which in general can be time, mass, and environment dependent.
The parameter $a_f$ {(which should not be confused with the scale factor $a$)} determines the radial dependence of the coupling in the screening limit along with $b$ that characterises an interpolation rate between the screened and unscreened limits.
Screening effects such as the chameleon
\citep{Li:2011qda,Khoury:2003aq,Lombriser:2013eza}
symmmetron \citep{Hinterbichler:2010es, Taddei:2013bsk}, k-mouflage \citep{Babichev:2009ee,  Brax:2014yla}, and Vainshtein \citep{Vainshtein:1972sx, Schmidt:2009yj, Dvali:2000hr} effects as well as other suppression effects such as the linear shielding mechanism~\citep{Lombriser:2014ira} or Yukawa suppression approximatively but analytically and sufficiently accurately map onto this transition function by specifying the expressions that the corresponding couplings assume in the limits of large and small $r$ and $r\rightarrow r_0$.
Furthermore, it was argued that in principle the parameter values can directly be determined from the action of a given gravitational theory after adopting the scaling method of \cite{McManus:2016kxu} and counting the powers of second and first spatial derivatives and the scalar field potential.

Note that one may also adopt other transition functions to interpolate between the two different regimes than Eq.~\eqref{eq:transition} such as a $\tanh$ or sigmoid~\citep{Hassani:2019wed} function and perform the analogous matching of the limits.

Finally, for our practical examples in Sec.~\ref{sec:test}, we will only consider models that recover GR in the fully screened limit, thus, $A=1$ in Eq.~\eqref{eq:nlparam}, and for which there is only one screening effect operating ($N_0=N_i=1$).
Hence, $B_i=B$ is given by the linear effective coupling in Eq.~\eqref{eq:GeffQS}, and $\mathcal{F}_{ij}=\mathcal{F}$ shall be specified by Eq.~\eqref{eq:transition}.

\subsection{Full parametrisation in Fourier space} \label{sec:general}

To model the effective modification of the gravitational coupling on all scales, the linear and nonlinear limits, Eqs.~\eqref{eq:GeffQS} and \eqref{eq:nlparam}, may be combined into one expression, for instance, by adopting Eq.~\eqref{eq:GeffQS} for $B_i$ in Eq.~\eqref{eq:nlparam}.
But this simple combination can only be performed for models where Eq.~\eqref{eq:GeffQS} is scale independent such as in nDGP gravity (Sec.~\ref{sec:nDGP}).
More generally, one must perform a Fourier transform of either Eq.~\eqref{eq:GeffQS} or Eq.~\eqref{eq:nlparam} to unify the effective modifications in either real or Fourier space. 
We discuss the advantages and disadvantages of the two different approaches in App.~\ref{Fourier_appendix}.
Due to the simplicity in solving Eq.~\eqref{eq:MGPoisson}, we adopt the Fourier space approach as our main method, and in the following we discuss the procedure we adopt to convert Eq.~\eqref{eq:nlparam} into a nonlinear effective gravitational coupling in Fourier space.

For this purpose, instead of the real space description in Eq.~\eqref{eq:nlparam}, we wish to write the parametrised gravitational coupling as
\begin{equation}
 \frac{\tilde{G}_{\rm eff}(a,k)}{G} = {A} + \sum_i^{{N}_0} B_i \prod_j^{{N}_i} \mathcal{\tilde{F}}_{ij} \,, \label{eq:Fourier_nlparam}
\end{equation}
where $\mathcal{\tilde{F}}_{ij}$ are now transition functions in Fourier space that parametrise screening or other suppression effects,
\begin{equation}
 \tilde{\mathcal{F} }\sim b \left(\frac{k_0}{k}\right)^{a_f} \left\{ \left[1+\left(\frac{k}{k_0}\right)^{a_f}\right]^{1/b} - 1 \right\} \label{eq:Fourier_transition}
\end{equation}
with $k_0$ characterising an effective screening Fourier wavenumber.
Note that Eq.~\eqref{eq:Fourier_nlparam} is not simply a recasting of Eq.~\eqref{eq:nlparam} into Fourier space, which instead involves complicated convolutions (see App.~\ref{Fourier_appendix}).

An immediate advantage of working in Fourier space is that we can now directly adopt
Eq.~\eqref{eq:GeffQS} for the linear limit $B_i$.
We stress, however, that the phenomenological parametrisation~\eqref{eq:Fourier_transition} can also be configured to match the scale dependence of the linear (or linearised) effective coupling given by Eq.~\eqref{eq:GeffQS}.
This is due to the applicability of the transition function to the Yukawa suppression \citep{Lombriser_2016}.
We test the performance of this description in Sec.~\ref{sec:linfR}.

The crucial aspect of converting Eq.~\eqref{eq:nlparam} into Eq.~\eqref{eq:Fourier_nlparam} is how the screening scale $r_0$ in Eq.~\eqref{eq:transition} must be reinterpreted for Eq.~\eqref{eq:Fourier_transition} in terms of $k_0$.
We adopt the following procedure for this conversion.
As in \cite{Lombriser_2016} for the computation of the modified spherical collapse, we first replace $r\rightarrow a \, \rth y_{\rm h}$, where $\rth$ is the comoving radius of a top-hat density $\rhom$ characterising a halo in our simulation and $y_{\rm h} = (\rhom/\brhom)^{-1/3}$ denotes the dimensionless top-hat radius with the cosmological background density $\brhom$.
We then perform the replacement $y_{\rm h}/y_0\rightarrow k_0/k$.
The dimensionless screening scale typically evolves in time and can also be dependent on mass and environment $y_{\rm env}$.
We thus perform the additional replacement $y_{\rm h}/y_{\rm env} \rightarrow k_{\rm env}/k$. Note that hereby the screening scale $k_{\rm 0}$ can hence become effectively $k$ dependent due to dependencies on mass and environment.

{It is worth emphasising that our conversion from the real to the Fourier space gravitational coupling can be motivated by the scaling method of \cite{McManus:2016kxu} (also see \cite{Lombriser_2016,McManus:2017itv,Renevey:2020tvr}), which isolates the dominant terms of a scalar field equation in the screened or unscreened regimes.
The radial dependence of the real-space $\Geff$ in these different limits is correctly obtained from replacing derivatives in the scalar field equation with the approximation $\nabla\rightarrow r^{-1}$.
Casting the scalar field equation into Fourier space, we convert derivatives as $\nabla^2\rightarrow-k^2$.
The replacement $y \sim r \sim k^{-1}$ can therefore be motivated based on scaling arguments.
}

For our simulations of parametrised modified gravity, we will focus on models with one screening transition with and without the addition of a linear Yukawa suppression.
In this case, the effective parametrised gravitational coupling of the nonlinear Poisson equation simplifies to
\begin{equation}\label{general_param}
 \frac{\tilde{G}_{\rm eff}(a,k) }{G} = 1 + 
 \frac{\Delta G_{\rm eff, L}}{G} b \left(\frac{k_0}{k}\right)^{a_f} \left\{ \left[1+\left(\frac{k}{k_0}\right)^{a_f}\right]^{1/b} - 1 \right\} \,, 
\end{equation}
where $\Delta G_{\rm eff}\equiv \Geff-G$.

As with the linear modification there are two options of using a parametrisation of $\Geff$.
One can either compute the functions $p_i(a)$ along with $k_0$, $b$ and $a_f$ for a specified modified gravity model, or one can phenomenologically parametrise these components and perform a simulation for a given set of parameter values.
Cosmological observations may then be used to generally constrain the available parameter space.
Here, we will follow the first approach and test the performance of the parametrised simulation framework with specific models against the simulation output of correspondent model-specific $N$-body codes.
We will, however, allow for a calibration of the model parameters in the expressions derived for $k_0$
or of the interpolation parameter $b$ against the model-specific simulations when not predicted analytically.
While the parameter values could be computed numerically by other means \citep{Lombriser_2016}, the motivation behind this approach is that in a parametrised approach it is primarily important that a set of parameter values can be chosen to reproduce a model. If observations favour a nonstandard set of parameter values, the exact model these parameters correspond to may still be determined in retrospect.
It is however generally feasible to replace the calibration with analytic predictions or simple numerical computations, and we expect future work to improve upon this point.
But this is beyond the scope of this first exploratory work.

\subsection{Implementation in $N$-body code} \label{sec:Nbody}

For our simulations of parametrised modified gravity, we use a Newtonian version of the \gev~$N$-body code \citep{Adamek:2015eda}, where we implement the parametrised gravitational coupling in Fourier space, Eq.~\eqref{general_param}.
We shall refer to this implementation as the \MG~code.
Note that \gev~is a particle-mesh $N$-body code, in which in the Newtonian version the Poisson equation is solved in Fourier space to update particle positions and momenta.
In contrast, in \MG~the modified Poisson equation is used to move the particles in the $N$-body code.
It is worth mentioning that \gev, as a relativistic $N$-body code, is mainly developed to study the relativistic effects in the cosmic large-scale structure.
Although we currently use the Newtonian version of \gev, it paves the way for an implementation of a modified gravity parametrisation that includes relativistic effects.

Due to the simplicity in our implementation, the run-time for a parametrised modified gravity simulation is almost the same as for $\Lambda$CDM runs.
This is a great advantage of the parametrised $N$-body code over exact model-specific simulations, which can slow down simulations tenfold \citep{Li_2012}.

To test the accuracy of our parametrised approach, in Sec.~\ref{sec:test} we compare our simulation output against that of other simulations for a range of well studied specific modified gravity models.

\section{Testing the $N$-body code} \label{sec:test}

In order to test our framework for the parametrised modified gravity $N$-body simulations introduced in Sec.~\ref{sec:MG} and its \MG~implementation in \gev~(Sec.~\ref{sec:Nbody}), we shall consider three toy scenarios: linearised $f(R)$ (Sec.~\ref{sec:linfR}), chameleon $f(R)$ (Sec.~\ref{sec:chamfR}), and nDGP gravity (Sec.~\ref{sec:nDGP}).
These are representative for the different types of suppression effects one can encounter in modified gravity models: linear effects (Yukawa) as well as screening by large potential wells (chameleon) and large derivatives (Vainshtein)~\citep{Joyce:2016vqv}.

For the \MG~simulations we use $N_{\rm pcl} = 256^3$ for the number of particles, a boxsize of $L = 200$ Mpc/h, and $N_{\rm grid} = 256^3$ for the number of grids.
These are relatively small simulations but suffice for our purpose.
The $f(R)$ and nDGP simulations used in this work for the comparison
are taken from~\cite{Cataneo:2018cic}, which were run using \texttt{ECOSMOG}~\citep{Li_2012}, an extension of the $\texttt{RAMSES}$ code~\citep{Teyssier_2002} that has been developed to simulate different classes of non-standard gravity models.
The cosmological parameters adopted in all of the simulations are $\Omega_{\mathrm{b}} h^{2}=0.02225$, $\Omega_{\mathrm{c}} h^{2}=0.1198$, $H_{0}=100 h=68~\mathrm{km}\:\mathrm{s}^{-1} \mathrm{Mpc}^{-1}$, $A_{\mathrm{s}}=2.085 \times 10^{-9}$ and $n_s = 0.9645$.
Our test quantity for the comparison of the output of the parametrised against the exact modified gravity model $N$-body simulations will be the matter power spectrum, for which we shall consider the range of scales where the upper bound in $k$, $2.5$~h/Mpc, is set according to the Nyquist frequency\footnote{We remove part of the data because of the error introduced by finite resolution effects.} of our \MG~simulations and the lower bound, $0.05$~h/Mpc, is set by their boxsize. {We note that since the data from the low-resolution simulations are very noisy, for a better comparison, we apply a smoothing with a Gaussian filter to remove the noise in our figures. Hereby a standard deviation of $\sigma = 2.5$ is adopted for the Gaussian kernels in all simulations. Specifically, we use the {\texttt gaussian\_filter} function in {\texttt SciPy~1.0}~\citep{2020SciPy-NMeth} to perform the smoothing.}

Like in \gev~in \MG~the initial conditions are configured using a linear Boltzmann code, here \class \citep{Lesgourgues:2011re}, at high redshifts, where perturbation theory is still valid.
We refer to App.~A of \cite{Adamek:2016zes} for more details on producing the high-redshift initial conditions with linear Boltzmann codes.
For the numerical results presented in this paper we use $z=100$ as the initial redshift.
Since the modifications of gravity of interest here reduce to GR at early times, we choose the same initial conditions for our modified gravity runs as for the $\Lambda$CDM simulation.
Note that we do not use the same seeds as used in \cite{Cataneo:2018cic}, but since we compare the relative difference between the modified gravity and $\Lambda$CDM matter power spectra, i.e. $\frac{P_{\text{MG}} - P_{\Lambda\text{CDM}}}{P_{\Lambda \text{CDM}}}$, the error introduced due to the cosmic variance is almost cancelled out.

\subsection{Linearised $f(R)$ gravity} \label{sec:linfR}

At linear scales of $f(R)$ gravity in the quasistatic limit, the modified Poisson equation~\eqref{eq:MGPoisson} in Fourier space takes the form
\be
k^2 \Phi_k = -4 \pi G \Big( \frac43 - \frac13 \frac{\mu^2 a^2}{k^2 + \mu^2 a^2} \Big) a \, \delta \rho_k \,, \label{lin_fR}
\ee
where $\Phi_k$ denotes the Fourier transform of the gravitational potential.
The Compton wavelength $\lambda$ of the scalaron field and its mass $\mu$ are specified by
\be
\mu^{-2} = \lambda^2 = \frac{-6 \bar{f}_{R0}}{3 H_0^2 (\Omega_m + 4 \Omega_{\Lambda})} \left(  \frac{1+ \frac{ 4  \Omega_{\Lambda}}{\Omega_m} }{a^{-3} + \frac{  4\Omega_{\Lambda}}{\Omega_m}}\right)^3 \,,
\ee 
where we have assumed a \cite{Hu:2007nk} model with exponent $\tilde{n}=1$.
The model parameter $\bar{f}_{R0} \equiv df(\bar{R})/d\bar{R}(z=0)$ parametrises the strength of the gravitational modification and together with the usual cosmological parameters fully specifies the $f(R)$ modification.
Eq.~\eqref{lin_fR} can be cast into Eq.~\eqref{eq:GeffQS} and can also be adopted at nonlinear scales, which corresponds to a linearisation of the effective gravitational coupling $\Geff$~\citep{Oyaizu:2008tb}.

Hence, while the mapping of linearised $f(R)$ gravity into our parametrised framework described by Eq.~\eqref{general_param} can be done exactly, we shall test here the performance of the parametrisation~\eqref{eq:Fourier_transition} for the Yukawa suppression described by Eq.~\eqref{lin_fR}.
Thus, we want to express Eq.~\eqref{lin_fR} as
\be  
\frac{\Delta G_{\rm eff}}{G}|_{\rm Yukawa} = \frac{b}{3} p \left[ \left(1+\frac{1}p\right)^{\frac{1}{b}} -1 \right] \label{Yukawa_param}
\ee
where $ p=\left( \frac{k b^n}{a \mu} \right)^{-a_{f}}$, $a_{f} = \frac{-2b}{b-1}$, and $n=\frac{b}{a_f(b-1)}$ can be inferred from the limits of Eq.~\eqref{lin_fR} following the procedure laid out in \cite{Lombriser_2016}.

\begin{figure*}
\begin{center}
\includegraphics[width=18cm]{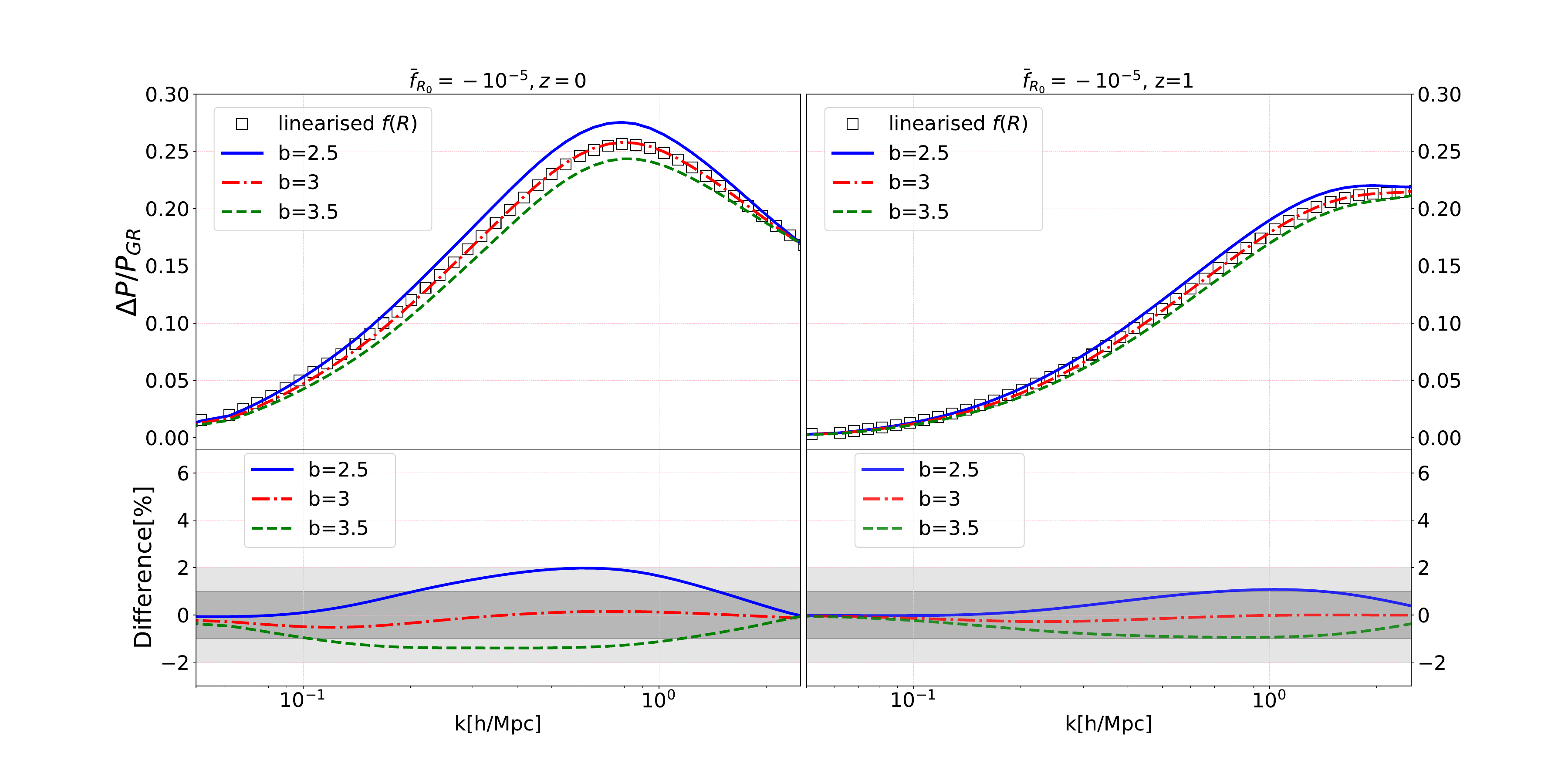} \vspace*{-6mm}\\
\includegraphics[width=18cm]{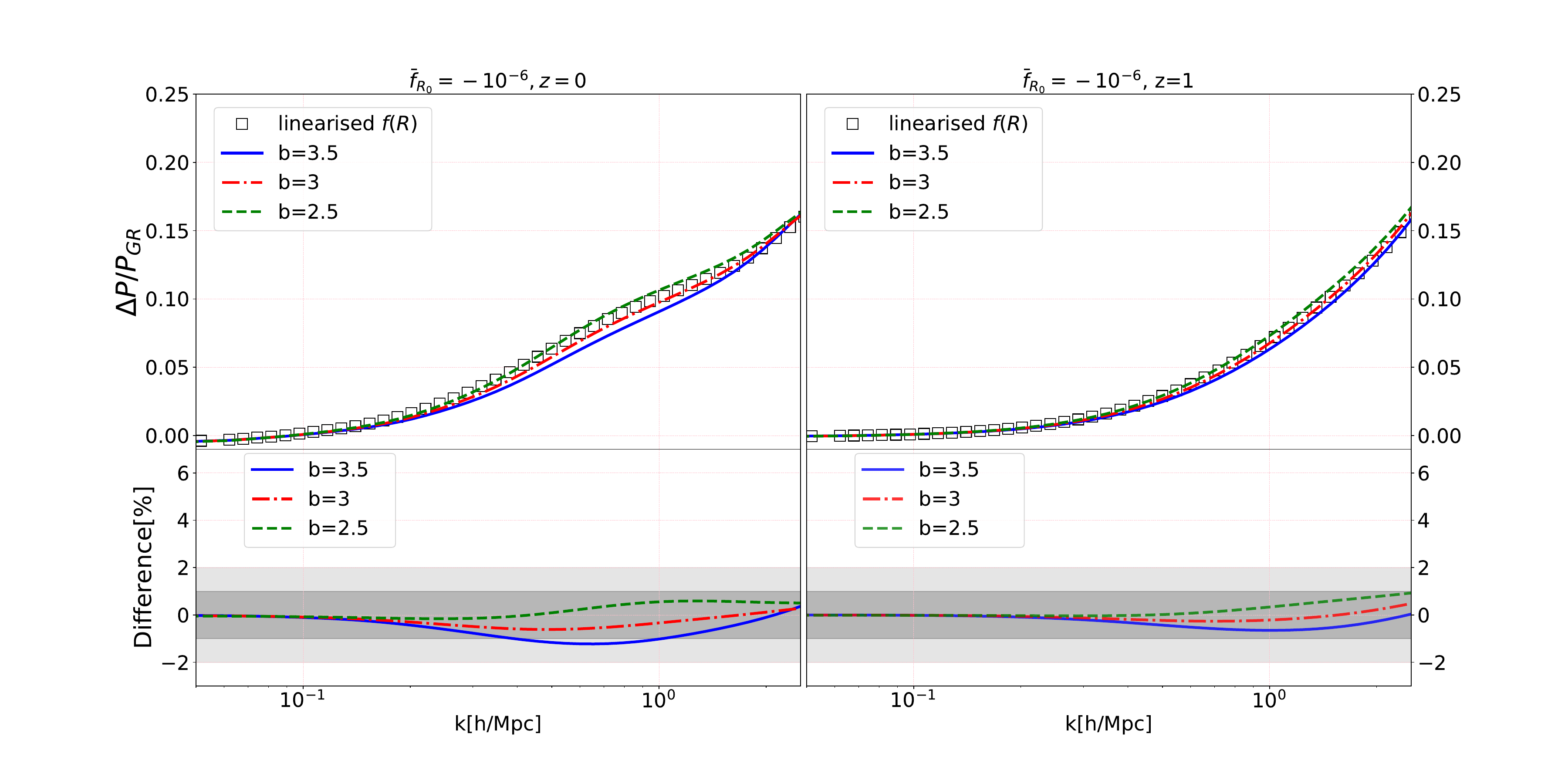}
\caption{Relative difference between the matter power spectra produced by the parametrised (Eq.~\eqref{Yukawa_param}), and exact (Eq.~\eqref{lin_fR}) \MG~implementations of the Yukawa suppression in linearised $f(R)$ gravity for redshifts $z=0$ (\emph{left panels}) and $z=1$ (\emph{right panels}) and three different values of the interpolation parameter $b$.
The simulations are run for $\bar{f}_{R0} = -10^{-5}$ (\emph{top panels}) and $\bar{f}_{R0} = -10^{-6}$ (\emph{bottom panels}).
The value $b=3$ provides a good match of the parametrised simulations to the exact implementation.
}
\label{fig:F5}
\label{fig:F6}
\end{center}
\end{figure*}

We compare the matter power spectra produced in the parametrised framework against those from the simulations of the exact modification, both implemented in \MG, in Fig.~\ref{fig:F5}.
We consider two strengths of the modification, $\bar{f}_{R0} =-10^{-5}$ and $\bar{f}_{R0} = -10^{-6}$, and two redshift slices at $z=0$ and $z=1$.
The parametrisation~\eqref{Yukawa_param}
produces an accurate match to the exact simulations with Eq.~\eqref{lin_fR} for all of these outputs and for the full range of scales up to $k=2.5$ h/Mpc.
The interpolation parameter seems to assume the universal value $b=3$ independent of redshift and strength of the modification.

While the parametrised transition function accurately reproduces the Yukawa suppression, as discussed in Sec.~\ref{sec:lin}, the adoption of Fourier space for the description of the effective gravitational coupling $\Geff(a,k)$ in the Poisson equation allows us to directly make use of the simple linear expression~\eqref{eq:GeffQS} instead.
We shall thus adopt Eq.~\eqref{eq:GeffQS} for the Yukawa regime in the following.

\subsection{Chameleon $f(R)$ gravity} \label{sec:chamfR}

\begin{figure*}
\begin{center}
\includegraphics[width=18cm]{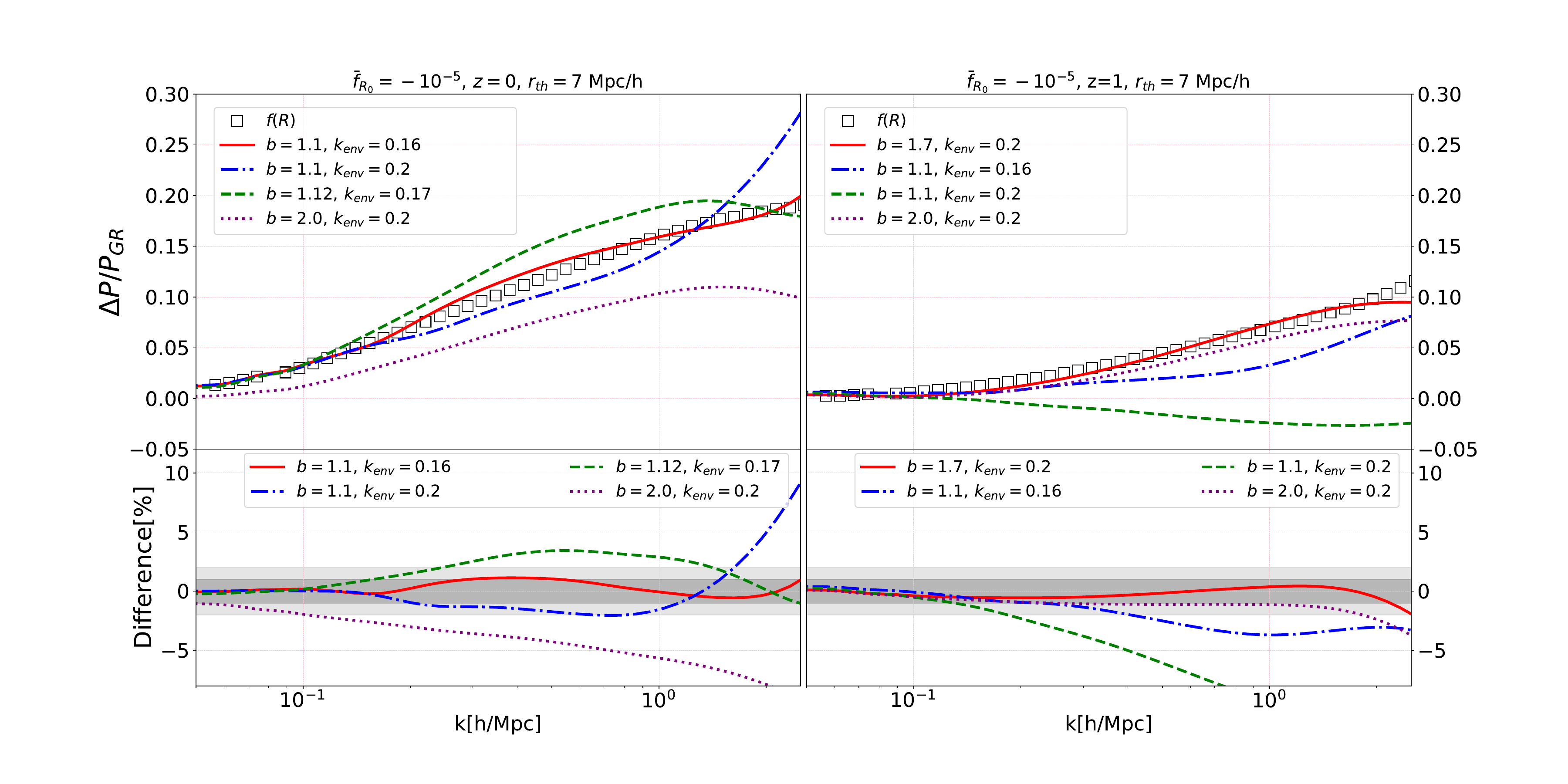} \vspace*{-6mm}\\
\includegraphics[width=18cm]{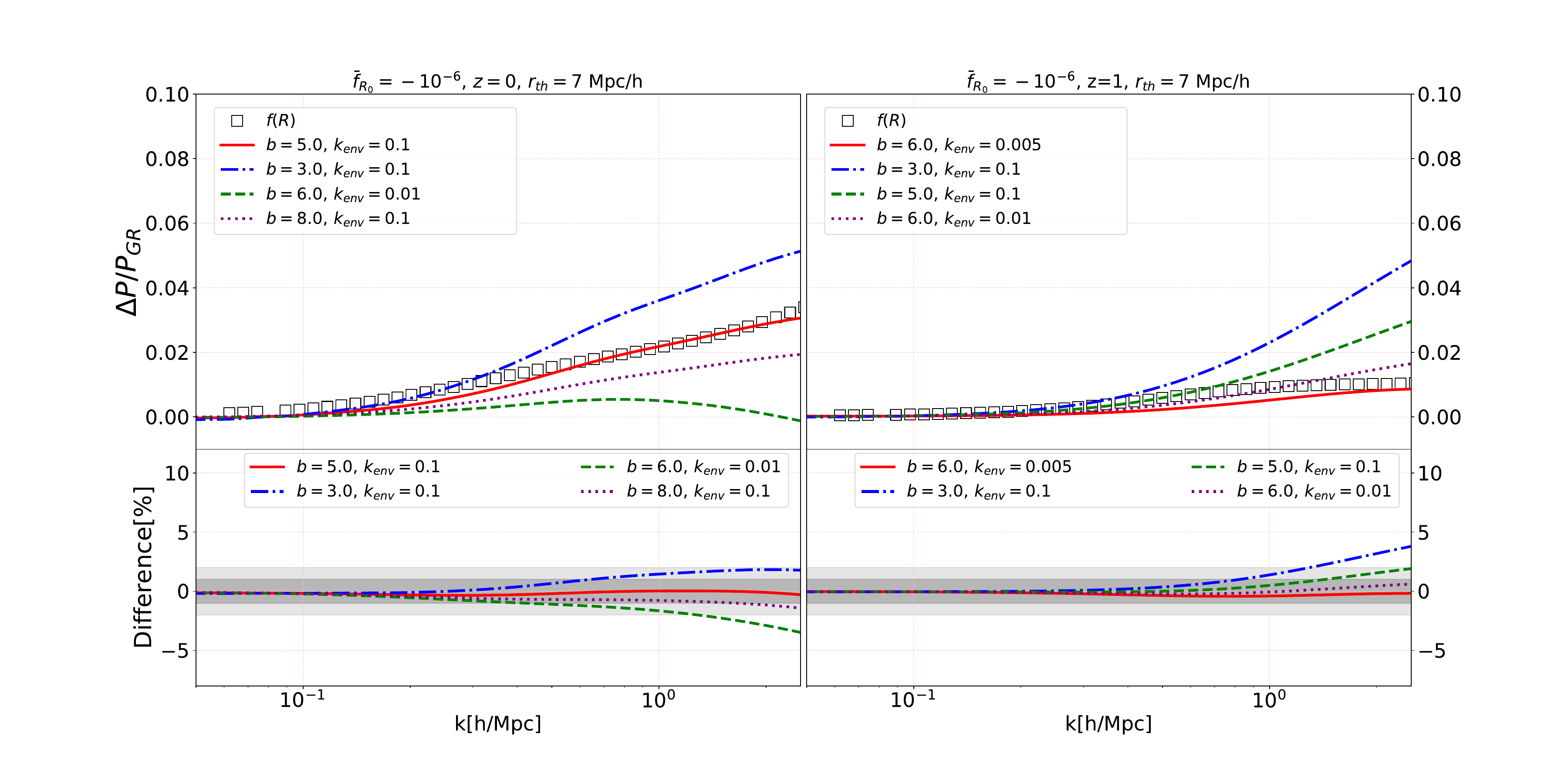}
\caption{
Same as Fig.~\ref{fig:F5} but for chameleon $f(R)$ gravity with comparison of the parametrised \MG~simulations against the exact model simulations
\citep{Cataneo:2018cic}.
The parametrised simulations were run for four different values each of the interpolation parameter $b$ and the environmental suppression scale $k_{\rm env}$, where the corresponding comoving top-hat radius was fixed to $r_{\rm th}=7$~Mpc/h.
\emph{Top panel}: For the values ($b =1.1$, $k_{\rm env} = 0.16~{\rm h/Mpc}$) at $z=0$ and ($b =1.7$, $k_{\rm env} = 0.2~{\rm h/Mpc}$) at $z=1$ we find a $\sim1\%$ match over all scales to $k = 2.5~{\rm h/Mpc}$.
\emph{Bottom panel}: A match of $\sim1\%$ is found for ($b =5.0$, $k_{\rm env} = 0.1~{\rm h/Mpc}$) at $z=0$ and ($b =6.0$, $k_{\rm env} = 0.005~{\rm h/Mpc}$) at $z=1$.
}
\label{fig:cham_fR5}
\label{fig:cham_fR6}
\end{center}
\end{figure*}

Next we shall consider parametrised simulations for the full $f(R)$ model, without the linearisation performed in Sec.~\ref{sec:linfR}.
We parametrise the effective gravitational coupling of the model with Eq.~\eqref{general_param}.
For the Yukawa regime, we can simply adopt Eq.~\eqref{eq:GeffQS} and hence we are left with a parametrisation of the chameleon screening mechanism by Eq.~\eqref{eq:Fourier_transition}.
More specifically, we write the parametrisation as
\be  
\frac{\Delta G_{\rm eff}}{G}|_{\rm tot} = \frac{\Delta G_{\rm eff}}{G}|_{\rm Yukawa}  \times  \frac{\Delta G_{\rm eff}}{G}|_{\rm Chameleon} \,,
\label{eq:cham_yukawa}
\ee
where $\frac{\Delta G}{G}|_\text{Yukawa}$ is given by Eq.~\eqref{lin_fR}.
The chameleon screening regime $\frac{\Delta G}{G}|_\text{Chameleon}$ in contrast is parametrised as
\be 
\frac{\Delta G}{G}|_\text{Chameleon} = b  \Big(\frac{k_{\rm 0}}{k}\Big)^{a_{f}} \; \Big\{ \Big[1+\Big(\frac{k}{k_{\rm 0}}\Big)^{a_f} \Big]^{\frac{1}{b}} -1 \Big\} \,.
\label{eq:chameleon_part}
\ee
To
find $\frac{k}{k_0}$ we will first inspect the real space parametrisation in Eq.~\eqref{eq:transition}, described by \cite{Lombriser_2016}.
Here, the screening scale $r_0$ is determined by the relation between thin-shell thickness $x$ of the chameleon mechanism and the physical top-hat radius $r_{\rm th}$.
More specifically,
\begin{equation}
 \frac{\Delta\Geff}{G} = \frac{1}{3} \left\{ 1 + \min\left[(x-1)^3,0\right] \right\}
\end{equation}
with thin-shell thickness
\begin{align}
 x = - C_1 r^7 \Big( C_2^{-2} -  C_3^{-2} \Big) \,, \label{eq_thinshell}
\end{align}
where the coefficients are given by
\begin{align}
 C_1 & = \frac{-\bar{f}_{R0}}{\Omega_m H_0^2 r_{\rm th}^3} \Big( \frac{\Omega_m + 4 \Omega_{\Lambda}}{4 \Omega_{\Lambda}} \Big)^2 \,, \\
 C_2 & = \frac{\Omega_m}{4  \Omega_{\Lambda} } r_{\rm th}^3 \,, \\
 C_3^2 & = C_2^2 \Big( \frac{y_{\rm h}}{y_{\rm env}} \Big)^6
\end{align}
and $y_{\rm h} = \frac{r}{a r_{\rm th}}$ is the dimensionless top-hat radius with $y_{\rm env}$ correspondingly characterising a dimensionless environmental radius.
We also have $y_0 = \frac{r_0}{a r_{\rm th}}$ for the dimensionless screening scale.
Note that $r$ is a physical radius whereas $r_{\rm th}$ is the comoving top-hat radius.

In the screened limit, we have $\Delta \Geff/G \approx x$ ($x\ll1$) and $\Delta \Geff/G \approx b(r/r_0)^7/3$ ($r \ll r_0$) in Eq.~\eqref{eq:transition}.
Performing the approximation $y\propto r \propto k^{-1}$, we obtain
\be 
\frac{r_0}{r} \rightarrow \frac{k}{k_0} = \Bigg\{ \frac{3 C_1}{b C_2^2} {\rm max} \bigg[ \Big( \frac{k}{k_{\rm env}} \Big)^6 -1 , 0 \bigg] \Bigg\}^{-1/7} \frac{k}{a} \,,
\ee
where
we have used the maximum function to
prevent negative screening scales when $k<k_{\rm env}$ with $k_{\rm env}$ denoting the effective environmental wavenumber.
There are three parameters in this expression, namely the interpolation rate $b$, the comoving top-hat radius $r_{\rm th}$ and the environmental Fourier wavenumber $k_{\rm env}$.
The top-hat radius and environmental wavenumber need to be understood here as effective, or average, quantities.
While $r_{\rm th}$ and $b$ are degenerate in $k/k_0$, $b$ also appears in $\frac{\Delta G}{G}|_\text{Chameleon}$.
In principle, these parameters could be determined from theory~\citep{Lombriser_2016}, but for the reasons discussed in Sec.~\ref{sec:general} we shall treat them as free parameters.
For simplicity, however, we set the comoving top-hat radius to $r_{\rm th} = 7$ Mpc/h, motivated by a typical galaxy cluster mass.

In Fig.~\ref{fig:cham_fR5} we compare the matter power spectra produced with our parametrised $N$-body simulations for the gravitational modifications in Eqs.~\eqref{Yukawa_param}, \eqref{eq:cham_yukawa} and \eqref{eq:chameleon_part} against the simulations of the exact $f(R)$ modification from \cite{Cataneo:2018cic} for $\bar{f}_{R0} = -10^{-5}$ and $\bar{f}_{R0} = -10^{-6}$ at redshifts $z=0$ and $z=1$.
We also vary $b$ and $k_{\rm env}$.
For $\bar{f}_{R0} = -10^{-5}$, we find that our parametrisation with values ($b = 1.1$, $k_{\rm env}$ = 0.16 h/Mpc) at $z=0$ and ($b = 1.7$, $k_{\rm env}$ = 0.2 h/Mpc) at $z=1$ provides a $\sim1\%$ level match to the simulated power spectra of the exact model over all scales to $k=2.5$ h/Mpc.
For $\bar{f}_{R0} = -10^{-6}$, we find a sub-percent level match for the parameters ($b=5.0$, $k _{\rm env} = 0.1$ h/Mpc) at $z=0$ and ($b=6.0$, $k _{\rm env} = 0.005$ h/Mpc) at $z=1$
over all scales to $k=2.5$ h/Mpc. %
It is worth noting that the match to the exact simulations could be improved by allowing for the additional variation of $r_{\rm th}$ or by a finer grid in the parameter space.
At this level of accuracy, however, one would also need to run higher-resolution simulations.

\subsection{Normal branch DGP gravity} \label{sec:nDGP}

\begin{figure*}
\begin{center}
\includegraphics[width=18cm]{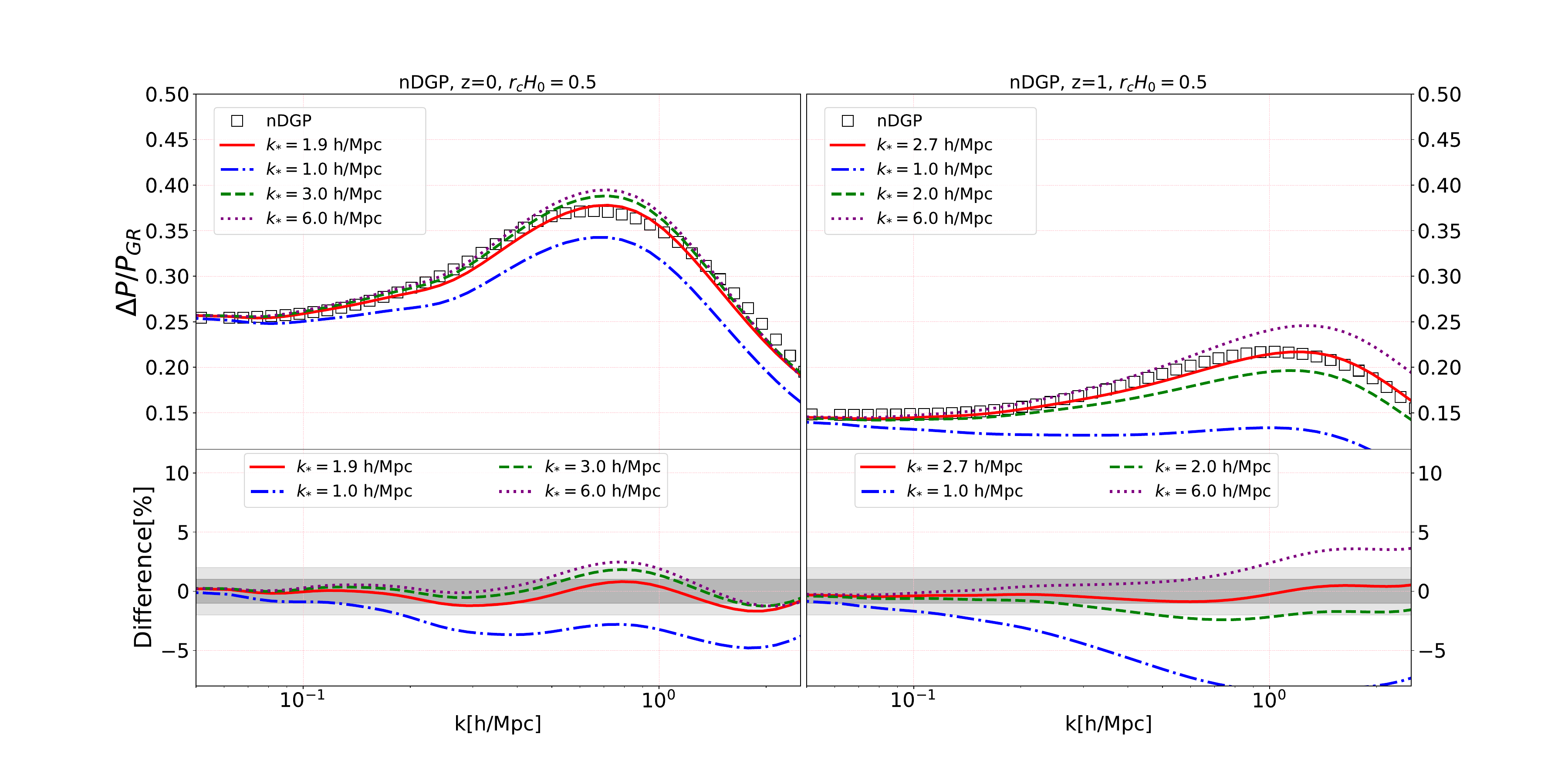}\vspace*{-6mm}\\
\includegraphics[width=18cm]{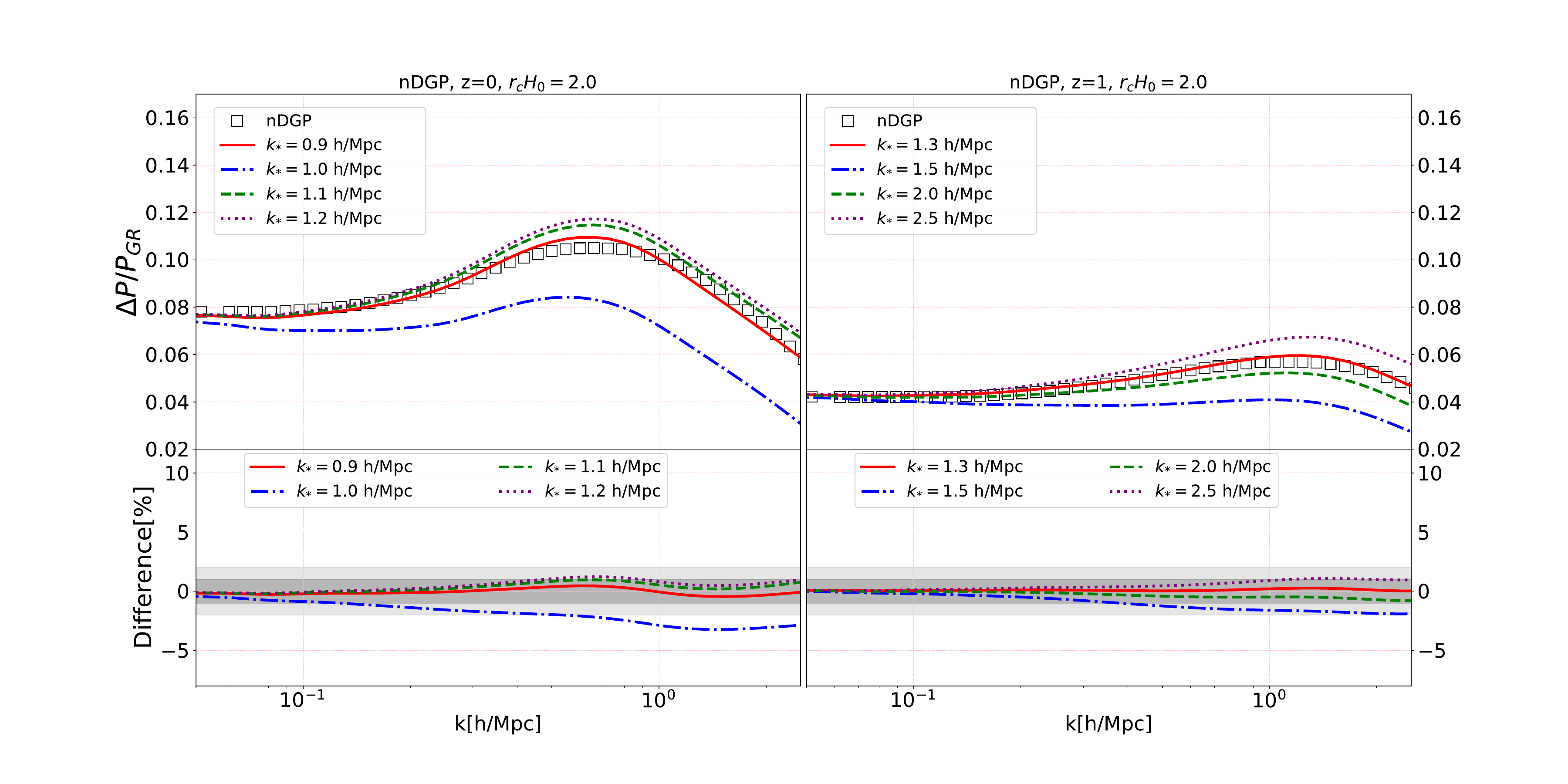}
\caption{Same as Fig.~\ref{fig:cham_fR5} but for nDGP gravity.
\emph{Top panels}: Comparison for the model parameter $H_0 r_c =0.5$ for four different values of the effective Vainshtein wavenumber $k_*$ at $z=0$ (\emph{left panel}) and $z=1$ (\emph{right panel}). The value $k_{\rm *} = 1.6$~h/Mpc at $z=0$ and $k_{\rm *} = 2.7$~h/Mpc at $z=1$ recovers the exact simulations at percent level across all scales to $k=2.5$~h/Mpc.
\emph{Bottom panels}: Comparison for $H_0 r_c =2.0$ for four different values of $k_*$ at  $z=0$ (\emph{left panel}) and $z=1$ (\emph{right panel}). The values $k_{\rm *} = 1.1~\rm{h}/\rm{Mpc}$ at $z=0$ and $k_{\rm *} = 2.0~ \rm{h}/\rm{Mpc}$ at $z=1$ provide sub-percent level matches.
}
\label{fig:nDGP_0d5}
\label{fig:nDGP_2}
\end{center}
\end{figure*}

Finally, we consider another widely studied modified gravity theory: the Dvali-Gabadadze-Porrati (DGP) braneworld model \citep{Dvali:2000hr,Deffayet:2000uy}.
For theoretical and observational consistency~\citep{Koyama_2005,Lombriser:2009xg}, we specify to the normal branch of the model (nDGP).
The new free parameter introduced here is the crossover scale $r_c$ that controls the leakage of gravity from the 4D brane to the 5D bulk spacetime.

In the linear limit of the nDGP model, or the weak-brane phase, the effective gravitational coupling $G_{\text{eff}}$ in the quasistatic modified Poisson equation~\eqref{eq:MGPoisson}
is scale independent and reads
\be
\frac{G_{\text{eff}}}{G}= 1 +  \frac{1} {3 \beta(a)} \,,
\ee
where the function $\beta(a)$ is defined as
\be
\beta(a) = 1+ \frac{4} {3 a }  \frac{\HH}{\HH_0} \HH_0 r_c   \Big ( 1+ \frac{\HH'}{2 \HH^2} \Big) 
\ee
with primes denoting derivatives with respect to the conformal time and $\HH$ indicating the Hubble expansion in conformal time.
More generally, due to the Vainshtein mechanism, caused by derivative self-interactions, this linear modification is suppressed in high-density regions, 
where the model recovers GR.
More specifically \citep{Koyama:2007ih,Schmidt:2009sg}, for a spherically symmetric matter density perturbation $\delta\rho$ we have
\be
\frac{G_{\text{eff}}}{G}= 1+ \frac{2}{3 \beta (a)}   \,\frac{ \sqrt{1+ x^{-3} }-1 }{x^{-3}} \label{nDGP_full}
\ee
where $x \equiv \frac{r}{r_*} $ and $r_*$ is the Vainshtein radius, 
\be
r_* = \Big( \frac{16 G \delta M r_c^2}{9 \beta^2} \Big) \label{rstar_eq}
\ee
with mass fluctuation
\be
\delta M= 4 \pi \delta \rho r^3/3 \,.
\ee
This can be rewritten as
\be
\frac{\Delta \Geff}{G} = \frac{2}{3 \beta} \frac{ \sqrt{1+\epsilon} -1} {\epsilon} \label{eq:GeffnDGP}
\ee
with
\be
\epsilon \equiv x^{-3} = \Big(\frac{r_*}{r}\Big)^3 = \frac{8 \HH_0^2 r_c^2}{9 \beta^2} \Omega_m(a) \delta =  \frac{8 \HH_0^2 r_c^2}{9 \beta^2} \Omega_{m,0} a^{-3} \delta \,.
\ee
Given that the form of the real-space transition in Eq.~\eqref{eq:transition} is motivated by nDGP it is trivial to cast Eq.~\eqref{eq:GeffnDGP} into
Eq.~\eqref{eq:transition}:
\be
B \to \frac{1}{3 \beta} , \;\; a_f \to 3 , \; \;b \to 2 , \;\; r_0 \to r_* \,.
\ee
We now wish to translate this into a parametrisation of the gravitational modification in Fourier space, Eq.~\eqref{eq:Fourier_transition}. Note, however, that we have also performed parametrised $N$-body simulations with this real-space expression and compared against the outputs of \cite{Cataneo:2018cic}, finding good agreement with those when applying a smoothing radius for $\delta$ (see App.~\ref{Fourier_appendix}). 

To obtain the parametrised modification in Fourier space, we perform the approximation
\be 
\epsilon = \Big(\frac{r_*}{r}\Big)^3 \to \Big(\frac{k}{k_0} \Big)^3 \,, \label{epsilon_fourier}
\ee
where $k_0=k_*$ is the wavenumber corresponding approximately to the Vainshtein radius $r_*$, and
we find
\be 
\frac{\Delta G}{G}|_\text{ \rm nDGP} = \frac{1}{3 \beta}  \Big(\frac{k_{\rm *}}{k}\Big)^3 \; \Big\{ \Big[1+\Big(\frac{k}{k_{\rm *}}\Big)^3\Big]^{\frac{1}{2}} -1 \Big\} \,.
\ee
While the effective screening wavenumber $k_*$ can in principle be modelled \citep{Lombriser_2016}, we shall treat it here as a free parameter following the discussion in Sec.~\ref{sec:general}.

As in the exact model simulations of \cite{Cataneo:2018cic}, we adopt a cosmological background that matches that of $\rm{\Lambda} CDM$, or equivalently we consider an artificial dark energy fluid that cancels out the effect of modified gravity in the background and as a result we obtain the same expansion history \citep{Schmidt:2009sg}.
We choose two different strengths of the modification, $H_0 r_c = 0.5$ and $H_0 r_c = 2.0$, for comparing with the exact simulations. 
Fig.~\ref{fig:nDGP_2} shows the matter power spectra produced in the $N$-body simulations of the parametrised and exact models for the two choices of $H_0 r_c$ at two redshifts, $z=0$ and $z=1$.
The parametrised simulations cover different values of the effective screening wavenumber $k_{\rm *}$.
For $H_0 r_c = 0.5$, percent matches are achieved for $k_* = 1.9$ h/Mpc at $z=0$ and $k_*=2.7$ h/Mpc at $z=1$ up to $k=2.5$ h/Mpc.
For $H_0 r_c = 2.0$ sub-percent level matches are found for $k_* = 0.9$~h/Mpc at $z=0$ and $k_* = 1.3$~h/Mpc at $z=1$.

\label{sec:maths}

\section{Conclusions} \label{sec:conclusions}
Einstein's Theory of General Relativity has been validated by an ever increasing amount of high-precision measurements ranging from the Solar System to micron scales.
However, its validation over cosmological distances at a comparable precision level remains an important endeavour.  
Additional motivation for cosmological tests of gravity is drawn from the requirement of a currently dominating dark energy contribution to explain the accelerated expansion of the Universe.
Over the next decade we will benefit from new cosmological surveys of unprecedented precision with which we will be able to put tight constraints on the cosmological properties of dark energy and modified gravity theories.
Of special interest will be the nonlinear regime of cosmic structure formation, where unique signatures are expected from the screening mechanisms that viable modified gravity theories must employ to recover GR in the well-tested Solar-System region. 
For robust predictions of the complex nonlinear structure, matching the observational precision with corresponding computational accuracy, we need to perform $N$-body simulations of the modified large-scale structure.
To date, an excessive amount of viable modified gravity theories
can be formulated based on the prospects of novel interactions of matter with new fields.
A systematic testing of the manifold cosmological implications from the possible modifications of gravity based on a model-by-model implementation in $N$-body codes is infeasible.

To overcome this limitation, in this paper we have proposed a parametrisation of the modified gravity effects on the linear and nonlinear cosmological structure formation adequate for $N$-body codes.
It is constructed from a parametrisation framework for linear theory and a parametrisation formalism for the deeply nonlinear scales, which is based on modified spherical collapse computations that incorporate the effects from the variety of available screening mechanisms.
Employing this framework, we have developed \MG, a Fourier-space implementation of this approach that is built on the Newtonian version of the \gev~$N$-body code.
We have tested our parametrised code with a number of widely studied modified gravity models,
including $f(R)$ and nDGP gravity, which encompass both large-field value and derivative screening effects with the employment of the chameleon and Vainshtein mechanisms, and for which exact $N$-body implementations are available. 
We have shown that the parametrised approach is capable of recovering the nolinear matter power spectra produced by the exact code implementations of these models to sub-percent accuracy up to the highly nonlinear scales of $k = 2.5$~h/Mpc covered by our simulations.

In future work we plan to explore our nonlinear parametrised gravity framework
employing higher resolution simulations that extend the results to larger wavenumbers.
We moreover wish to apply and test our parametrised code with further modified gravity models.
Finally, we envisage the employment of the code for observational applications, offering an accurate generalised modelling tool for the exploitation of nonlinear data from forthcoming cosmological surveys in large-scale tests of gravity.

\section*{Acknowledgements}
We thank Matteo Cataneo and Baojiu Li for providing $f(R)$ and nDGP simulation outputs.
FH would like to thank Mona Jalilvand for assistance with the numerical simulations and Alessandro Casalino for correcting couple of typos in the paper. 
FH acknowledges support by Project Funding of the Swiss National Science Foundation (SNSF) (No.~182231).
LL was supported by a SNSF Professorship grant (No.~170547). 
Numerical computations were performed on the Baobab cluster of the University of Geneva.
\section*{Data availability}
The
code developed for the simulations, data and figures of
this paper will be shared on request to the corresponding author.

\bibliographystyle{mnras}
\bibliography{Ref}

\appendix

\section{Semi-dynamical perturbations} \label{app:semi_analytics}

In Sec.~\ref{sec:lin} we have adopted the quasistatic limit, neglecting time derivatives over spatial deriviatives, for our description of the effective gravitational coupling $\Geff$ in the modified Poisson equation.
While this approximation is accurate in linear theory of Horndeski scalar-tensor modifications of gravity at scales well below the sound horizon \citep{Lombriser_2015}, it can break down at near horizon scales and beyond.
More accurately, one may therefore describe the modification of the Poisson equation within a semi-dynamical approximation, where time derivatives are evaluated and included at a pivot scale.

Following \cite{Lombriser_2015} we obtain for $\Geff = \mu(a,k)$ the $k$-dependent expression
\begin{equation}
 \Geff(a,k) = \frac{1}{8\pi M^2} \frac{\mu_{+2} k_H^2 + \mu_{+4} k_H^4 + \mu_{+6} k_H^6}{\mu_{-0} + \mu_{-2} k_H^2 + \mu_{-4} k_H^4 + \mu_{-6} k_H^6} \,, \label{eq:Gefflin}
\end{equation}
where $\mu_{\pm i}$ and $M^2$ are functions of time only, specified in \cite{Lombriser_2015}, and $k_H \equiv k/(aH)$.
Note that in the small-scale limit, formally where $k\rightarrow\infty$, this simplifies to
\begin{equation}
 G_{{\rm eff},\infty}(a) = \frac{\mu_{+6}}{\mu_{-6}} = \frac{1}{8\pi M^2} \frac{\mu_{\infty}^+ + \alpha_{\rm H} \left( f_{\Psi} \mu_{\Psi,\infty}^+ + \fzeta \mu_{\zeta,\infty}^+ \right)}{\mu_{\infty}^- + \alpha_{\rm H} \left( f_{\Psi} \mu_{\Psi,\infty}^- + \fzeta \mu_{\zeta,\infty}^- \right)} \,, \label{eq:GeffSD}
\end{equation}
where $f_{\Psi}\equiv d\ln\Psi/d\ln a$ and $f_{\zeta}\equiv d\ln\zeta/d\ln a$ encapsulate time derivatives of the perturbations with $\zeta$ denoting the comoving curvature.
The functions $\mu_{\infty}^{\pm}$ and $\mu_{\Psi\wedge\zeta,\infty}^{\pm}$ are time dependent but independent of $f_{\Psi}$ and $f_{\zeta}$ such that velocity fields and time derivatives of the spatial metric potential only contribute for beyond-Horndeski models ($\alpha_{\rm H}\neq0$) at leading order in the small-scale limit \citep{Lombriser_2015}.
Hence, for Horndeski theories, at leading order one can set $f_{\zeta}=f_{\Psi}=0$, in which case $\Geff(a,k)$ can directly be expressed by the time-dependent EFT functions $\{\alpha_i\}$, $\{\Omega,\Gamma,\Lambda,\ldots\}$, or the inherently stable basis~\citep{Kennedy:2018gtx,Lombriser:2018olq}.

Finally, alternatively to adopting the expression~\eqref{eq:Gefflin}, one may simply compute the linear perturbations and evaluate $\Geff$ from the modified Poisson equation~\eqref{eq:MGPoisson}.

\section{Fourier versus real space simulations} \label{Fourier_appendix}

As discussed in Sec.~\ref{sec:nl}, while the linear modification $B$ in Eq.~\eqref{eq:nlparam} is specified in Fourier space, the nonlinear expression for $\mathcal{F}$ is generally given in real space.
It is in general computationally not feasible to find the nonlinear modification to gravity in Fourier space from the modification in real space (e.g., nDGP) as one has to deal with convolutions.
The Fourier transformation $\mathcal{F_T}$
 of the real-space Poisson equation,
\be 
\mathcal{F_T} \big\{ \nabla^2 \Phi_N \big\} = \mathcal{F_T} 
\big\{ (1 + \frac{\Delta \Geff}{G} ) \delta \rho \big\} \,,
\ee
using the convolution theorem, gives
\be
-k^2 \Phi_N = \mathcal{F_T} \big\{1 + \frac{\Delta G_{eff}}{G} \big \} * \mathcal{F_T} \big\{\delta \rho \big\} \,,
\ee
where $*$ refers to the convolution.
Calculating this expression analytically is not practical since we do not have access to the full $\delta \rho$ analytically and moreover, computing it numerically is also not feasible because we need to integrate over all the lattice points in Fourier space which is contradictory to the nature of $N$-body simulations as the equations are solved in parallel and each part of the lattice only has access to its neighbourhood.
Due to these complications we have therefore resorted to the effective parametrised Poisson equation~\eqref{general_param}.

It is however worth mentioning that we nevertheless performed real-space simulations for the parametrisations of the $f(R)$ and nDGP cases.
In these simulations, while we found satisfactory results for the small scale behaviours, i.e., the chameleon and Vainshtein screening limits, for the large-scale behaviours we initially did not.
To overcome the issues at large scales we used a smoothed density field instead of the local one in the real-space parametrised modified Poisson equation.
We performed the smoothing with a Gaussian window function $W_{G}(\vec x; R)$ defined as
\be
W_{G}(\vec x; R)  = \frac{3}{4\pi R^3} e^{-|x|^2/2 R^2} \,,
\ee
where $R$ is the smoothing radius and we can construct the smoothed density field ${\delta}(\vec x; R)$ from the local value $\delta(\vec x)$ through
\be
\delta(\vec x; R) = \int \delta(\vec{x}')W_{G}(\vec x; R) d\vec x' \,.
\ee
We employed the convolution theorem to simplify this expression to
\be
 \hat \delta(\vec k; R) = \delta(\vec k) W_{G}(\vec k; R)
\ee
such that finally, the density smoothed over the radius $R$ with a Gaussian window function reads
\be
 \delta(\vec x,R) = \mathcal {F_T}^{-1} \Big[ \hat{\delta} (\vec k,R)  e^{-(k R)^2/2}  \Big] \,.
\ee
This procedure produces relatively consistent results for the parametrised real-space nDGP simulations on all scales.
The results for the large-scale behaviour in the parametrised real-space $f(R)$ simulations, however, does not improve.
This can be attributed to the scale dependence in the Yukawa suppressed regime of $f(R)$ gravity, which is not present for the nDGP modification.
Having found good agreement of the parametrised Fourier-space simulations with the exact simulations in Sec.~\ref{sec:test}, we leave an improvement of the parametrisation of the Yukawa regime in real space to future work.

\label{lastpage}
\end{document}